# Maximizing User Experience with LLMOps-Driven Personalized Recommendation Systems

Chenxi Shi[1*], Penghao Liang[1], Yichao Wu[2], Tong Zhan[3], Zhengyu Jin[4]

[1*] Software development, Telecommunication Systems Management, Northeastern University, Boston, MA, USA
[1] Information Systems, Northeastern University, Boston, MA, USA
[2] Computer Science, Northeastern University, Boston, MA, USA
[3] Computer Science, Columbia University, NY, USA
[4] Informatics, Univeristy of California, Irvine, Irvine, CA, USA
*Corresponding author: Chenxi Shi[E-mail:chxishi@gmail.com]

## ABSTRACT

The integration of LLMOps into personalized recommendation systems marks a significant advancement in managing LLM-driven applications. This innovation presents both opportunities and challenges for enterprises, requiring specialized teams to navigate the complexity of engineering technology while prioritizing data security and model interpretability. By leveraging LLMOps, enterprises can enhance the efficiency and reliability of large-scale machine learning models, driving personalized recommendations aligned with user preferences. Despite ethical considerations, LLMOps is poised for widespread adoption, promising more efficient and secure machine learning services that elevate user experience and shape the future of personalized recommendation systems.

## 1. INTRODUCTION

Personalized recommendation methods, whether based on machine learning or deep learning, have been central to research efforts aimed at improving user experience. These methods leverage various techniques to analyze user behavior and preferences, ultimately providing tailored recommendations. In today's digital era, personalized recommendation systems play a vital role in enhancing user satisfaction and engagement across a wide range of platforms and services. These systems offer customized content to users by analyzing their behavior, preferences, and historical data. For example, platforms like Netflix and Amazon utilize personalized recommendation systems to suggest movies, TV shows, and products based on individual user data. This personalized approach not only facilitates content discovery but also increases user satisfaction and loyalty to the platform.

One of the key benefits of personalized recommendation systems is their ability to significantly reduce the time users spend searching for relevant information or products. By presenting users with personalized recommendations, these systems streamline the decision-making process, saving users time and effort. Moreover, personalized recommendations promote user interaction and stickiness, fostering long-term engagement and loyalty.

To further enhance the effectiveness of personalized recommendation systems, prompt engineering technology can be integrated. Prompt engineering involves adjusting input prompts to guide model output, ensuring that recommendations are consistent and aligned with user preferences. By leveraging prompt engineering techniques, personalized recommendation systems can deliver more accurate and relevant recommendations, thereby improving overall user experience.

In conclusion, personalized recommendation systems are essential for improving user experience in today's digital landscape. By analyzing user data and offering tailored recommendations, these systems not only enhance user satisfaction and engagement but also promote long-term user loyalty and platform growth. Through the integration of prompt engineering technology, personalized recommendation systems can further optimize recommendations, ensuring a seamless and personalized user experience.

## 2. RELATED WORK

### 2.1 Development and challenge of personalized recommendation system

The recommendation system deduces user preferences by obtaining user's historical behavior data, such as web browsing data, purchase records, social network information, and user's geographical location. With the development of computer technology, the recommendation technology adopted by the recommendation system is mainly based on the early user-item-based data matrix decomposition technology, and gradually develops in the direction of integrating with data mining, machine learning, artificial intelligence and other technologies, so as to deeply explore the potential preferences of users' behaviors and build a more accurate user preference model. Zhang and Jiao (2007) proposed a recommendation system based on association classification, aiming at personalized recommendation services in B2C e-commerce applications. Their research combines the concepts of relational classification and personalized recommendation to achieve targeted recommendations by analyzing users' historical purchasing behavior and product attributes. This method provides an effective recommendation mechanism for e-commerce platform, which can improve users' shopping experience and platform sales efficiency.

In their research, Naumov et al. (2019) propose a deep learn-based personalized recommendation model that aims to improve the effectiveness of personalization and recommendation systems. The model utilizes deep learning techniques to enable more accurate and personalized recommendations by learning the complex relationships between users and items. Their research provides new ideas and methods for the development of recommendation system, and promotes the progress and application of personalized recommendation technology. Through the introduction of deep learning-driven recommendation models, Naumov et al. (2019) made an important contribution to the development and current status of personalized recommendation systems. Their research highlights the potential of deep learning in recommendation systems, opening up new possibilities for improving user experience and recommendation accuracy. This study provides useful guidance and enlightenment for the future development direction of personalized recommendation system.

Therefore, in the study of personalized recommendation systems, the work of Zhang and Jiao (2007) emphasized the potential of association classification methods in e-commerce applications. By combining association classification technology with recommendation system, they provide a new idea and solution for personalized recommendation. This research is of great significance to the development of personalized recommendation technology in the field of e-commerce, and provides beneficial reference and enlightenment for realizing more accurate and effective personalized recommendation.

### 2.2 User experience optimization

In large software organizations, the challenges and implications of user experience optimization extend beyond the lack of customer data sharing. The research of Fabijan et al. (2016) also reveals other aspects of the problem, such as:

1. Limitations of personalized recommendation: Lack of customer data sharing may lead to performance degradation of personalized recommendation systems. Due to lack of access to sufficient user behavior data, personalized recommendation algorithms may not accurately understand user preferences and needs, thus affecting the quality and accuracy of recommendations.

2. Reduced user engagement: If users feel that their data is not properly protected and used, they may have doubts about participating in the use of software products. This can lead to decreased user engagement with the product, which in turn affects the optimization of the user experience and the development of the product.

3. Trust and privacy issues: The lack of customer data sharing can raise concerns about the privacy and security of personal data. If users cannot trust the software organization to properly handle their personal data, they may choose not to share the data with the organization, which will further limit the development and optimization of personalized recommendation systems.

In summary, the lack of customer data sharing has a profound impact on user experience optimization, not only affecting the performance of personalized recommendation systems, but also potentially leading to reduced user engagement and increased trust issues. Therefore, large software organizations should attach importance to customer data sharing and take effective measures to solve this problem to promote the continuous optimization of user experience and the long-term development of the organization.

Personalized recommendation system has shown vigorous development in the aspect of diversified recommendation algorithms. Among them, the IBM.com personalized user experience work studied by Karat et al. (2003) explored the application of collaborative filtering, content-based recommendation and other algorithms in personalized recommendation. This work highlights the advantages and applicable scenarios of different algorithms, and provides valuable experience and guidance for the algorithm selection of personalized recommendation system. Data-driven is the core of personalized recommendation system, and the personalized user experience of IBM.com studied by Karat et al. (2003) provides strong support for this. Through a large number of user behavior data and item information, personalized recommendation system can achieve accurate recommendation services. The research of Karat et al. highlights the importance of data in recommendation algorithms, laying a solid foundation for the development and performance improvement of personalized recommendation systems.

## 2.3 LLMOps-Driven Personalized Recommendation Systems

According to the study of Kulkarni et al. (2023), the application of LLMOps (Large Language Model Operations) in enterprises has attracted wide attention. LLMOps refers to the operation and strategy of managing, optimizing and extending LLM (Large Language Model), which is of great significance for the development of personalized recommendation system. LLMOps can improve the efficiency and performance of the model to provide more accurate and faster recommendation services for personalized recommendation systems, thereby optimizing the user experience. At present, the application of LLMOps in personalized recommendation systems is in the initial stage of development. With the progress of large-scale language model technology and the continuous evolution of LLMOps, people begin to realize the potential of LLMOps in personalized recommendation system. Through the personalized recommendation system driven by LLMOps, the ability of large-scale language models can be better utilized to provide users with more personalized and accurate recommendations, thus improving the user experience and the competitiveness of the platform.

Despite the promising prospects of LLMOps, there remain several challenges and limitations to address. These include constraints related to model size and computational resources, as well as considerations regarding data privacy and security. Additionally, there is a need for ongoing algorithm and model optimization to fully realize the potential of LLMOps in personalized recommendation systems.Integrating prompt engineering technology into LLMOps-driven recommendation systems offers several advantages. Prompt engineering enables the customization of input prompts to guide model output, ensuring that recommendations are aligned with user preferences. By incorporating prompt engineering techniques, LLMOps-driven recommendation systems can further enhance recommendation accuracy and relevance, ultimately providing users with a more intelligent and personalized recommendation experience.

Looking ahead, as technology continues to advance and LLMOps matures, LLMOps-driven personalized recommendation systems are poised to become a significant development direction in the field of personalized recommendation. By addressing existing challenges and leveraging prompt engineering alongside LLMOps, these systems can deliver more tailored and impactful recommendations, further improving the user experience.However, despite the promising application of LLMOps in personalized recommendation systems, there are still some challenges and limitations. These include model size and computational resource constraints, data privacy and security considerations, and the need for algorithm and model optimization. In the future, with the continuous progress of technology and the maturity of LLMOps, it is believed that LLMops-driven personalized recommendation systems will become an important development direction in the field of personalized recommendation, and bring users a more intelligent and personalized recommendation experience.

1)LLMOps implementation steps

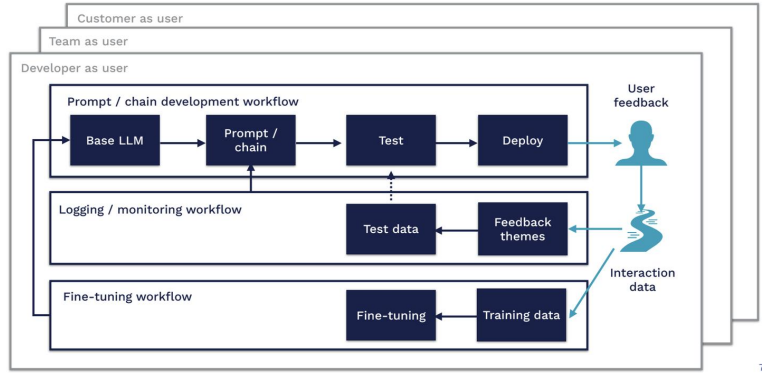

**Figure 1.** Step-by-step process of LLMops

One of the issues that LLM mentioned in its production survey was model accuracy and hallucinations.

This means that getting output from the LLM API in the format you want can take some iteration, and, if LLMS don't have the specific knowledge they need, they can hallucinate. To address these issues, you can adapt the base model to downstream tasks by:

2) Prompt Engineering: is a technique of adjusting the input to make the output conform to your expectations. You can use different techniques to improve your Prompt. One way is to provide some examples of the expected output format. This is similar to zero-sample or small-sample learning Settings. Tools such as LangChain or HoneyHive have emerged to help you manage and version prompt templates.

3) Fine-tuning pre-trained models is a known technique in ML. It can help improve the performance of the model on specific tasks. Although this increases the training effort, it reduces the inference cost. The cost of the LLM API depends on the input and output sequence length. Therefore, reducing the number of tokens you enter reduces API costs because you no longer have to provide an example in the prompt.

4) External Data: The underlying model often lacks contextual information (e.g., access to some specific document or email) and can quickly become obsolete (e.g., GPT-4 was trained on data until September 2021). Because LLMS can hallucinate if they don't have enough information, we need to be able to give them access to relevant external data. There are already tools available, such as LlamaIndex (GPT Index), LangChain, or DUST, that can act as a central interface to connect (" link ") LLMS to other agents and external data.

5) Embeddings: Another approach is to extract information from the LLM API in the form of embeddings (for example, movie summaries or product descriptions) and build applications on top of them (for example, search, compare, or recommend). If np.array is not enough to store your long-term memory embedding, you can use vector databases such as Pinecone, Weaviate, or Milvus.

## 3. METHODOLOGY

### 3.1 Prompt engineering technology

Prompt engineering refers to adjusting the input prompts to guide the model to produce the desired output when using the AI model. In the field of artificial intelligence, especially in tasks such as natural language processing and image generation, Prompt engineering is becoming increasingly important. By designing the appropriate Prompt, you can effectively influence the output of the model, making it more in line with expectations, and improving the interpretability and controllability of the model. Prompt engineering can help the model better understand the user's intentions, reduce misunderstandings and error outputs, and thus improve the system's performance and user experience.

A variety of Prompt engineering techniques can be used for different tasks and scenarios. A common technique in natural language processing tasks is to use sample output to guide the model to generate a specific type of text. For example, in a language translation task, a pair of sentences can be provided as input prompts to guide the model to produce the corresponding translation result. Another common technique is to use control tokens to adjust the generation behavior of the model, such as adding specific tags to a text generation task to indicate the style or theme of the generation. In image generation tasks, Prompt engineering techniques can direct the model to generate images of a particular style, content, or composition by adjusting the input visual description or adding specific markup. In conclusion, the choice of the appropriate Prompt engineering technology depends on the specific task requirements and the characteristics of the model, through proper Prompt design, you can improve the performance and scope of the model, and achieve a more intelligent and personalized output.

2) Application of Prompt engineering technology in personalized recommendation system

1. In the field of personalized recommendation, Prompt engineering technology can improve the accuracy and user experience of the recommendation system. For example, consider an e-commerce platform that wants to recommend personalized items to users. With Prompt engineering, you can design A series of input prompts that are specific to user preferences, such as "Based on your recent purchase of X, we recommend you browse Y" or "Your favorite category is Z, and we recommend related product A." By analyzing a user's purchase history, browsing behavior, and preferences, as well as actual data related to those behaviors, you can generate personalized Prompt and guide the recommendation system to produce recommendations that are more accurate and aligned with the user's interests. In this way, users will get a more personalized recommendation experience, increasing user satisfaction and loyalty to the platform.

2. Prompt engineering can also play an important role in user experience. Taking intelligent voice assistant as an example, Prompt engineering can improve the response accuracy and user interaction experience of voice assistant. By designing input prompts with clear semantics, such as "Please tell me what you want to do" or "What help do you need," the voice assistant can be guided to better understand the user's intentions and provide accurate responses. By analyzing user interaction data and actual conversation records with the voice assistant, the design of input prompts can be optimized and a more intelligent and personalized interactive experience can be provided. In this way, users will be more willing to use voice assistants and feel satisfied with their interactive experience, improving user recognition and frequency of use of products.

### 3.2 Introduction to the application of LLMOps in personalized recommendation system

LLMOps, or Large Language Model Operations, refers to the operation and strategy of managing, optimizing, and extending Large Language Models (LLMs) within personalized recommendation systems. Its role is pivotal, as LLMOps plays a crucial role in enhancing the efficiency and performance of LLMs, ultimately improving the accuracy and relevance of recommendations provided by personalized recommendation systems. Significantly, LLMOps enables the seamless integration and deployment of LLMs within recommendation systems, ensuring that these models operate effectively and deliver personalized recommendations tailored to individual user preferences and behavior. In essence, LLMOps serves as the backbone of personalized recommendation systems, driving their effectiveness and ensuring their ability to meet user demands and expectations in today's dynamic digital landscape.

The working principle of LLMOps revolves around managing, optimizing, and extending Large Language Models to enhance the performance of personalized recommendation systems. By employing advanced techniques such as model fine-tuning, resource allocation, and optimization of computational processes, LLMOps ensures that LLMs operate efficiently and effectively within recommendation systems. This optimization process involves fine-tuning the model parameters based on real-time user interactions and feedback, allocating computational resources to prioritize tasks that contribute to recommendation accuracy, and extending the capabilities of LLMs through continuous training and updates. The impact of LLMOps on recommended system performance is profound, as it leads to improved recommendation accuracy, reduced latency in generating recommendations, and enhanced user satisfaction and engagement. Ultimately, LLMOps empowers personalized recommendation systems to deliver more accurate, relevant, and timely recommendations, thereby optimizing the overall user experience and driving business success.

### 3.3 Comparative analysis of LLMOps and MLOps

1) Data pipeline

LLMOps emphasizes trust in data sources and interpretability of data, and tends to capture less but higher quality data. In the case of e-commerce, LLMOps may choose to use only sales data from reliable suppliers and ensure that this data is accurate and complete. Its data management process includes rigorous screening and evaluation of data sources, possibly vendor reputation investigations and verification to ensure the credibility of the data. Subsequently, the data is cleaned and preprocessed to eliminate any outliers and missing values and ensure data quality. Finally, through data annotation and classification, such as classifying sales data according to product category, sales region, etc., in order to better apply to the training and testing of recommendation systems or predictive models.

In contrast, MLOps focuses more on the diversity and quantity of data and on improving data quality through data engineering and data science approaches. In e-commerce, MLOps may collect data from different channels, including sales data, user behavior data, marketing data, etc., and build sophisticated data warehouses and data lakes to store and integrate this data. Then, data mining and machine learning technologies are used to conduct in-depth analysis and feature engineering of the data, such as extracting user preference characteristics through user behavior data, so as to achieve personalized recommendation. Finally, model training and tuning techniques are used to improve the effect and performance of the model. For example, collaborative filtering algorithms or deep learning models are used to train the recommendation system to improve the recommendation accuracy and user satisfaction.

To sum up, there are some differences between LLMOps and MLOps in data management processes. LLMOps focuses on data quality and interpretability, and ensures data accuracy and reliability by strictly screening and cleaning data. MLOps pays more attention to the diversity and quantity of data, and improves data quality through data engineering and data science methods, and is applied in model training and optimization to achieve better business results and user experience.

2) LLMOps model experiment process

LLMOps emphasizes the normativity and reproducibility of experiments, requiring detailed recording and interpretation of model experiments, including experimental steps, data sets, model parameters, etc. Experimental design and execution are required, as well as analysis and interpretation of experimental results. In addition, interpretability analysis of the model is also required to better understand the operating mechanism of the model.

The design of model experiments under LLMOps focuses on specification and reproducibility. For example, in an e-commerce recommendation system, an experimental design may be to test the performance of different recommendation algorithms (such as collaborative filtering, content filtering) on the same data set.

Performing experiments requires careful documentation of each step, including data preprocessing, model training, validation, and testing. What is recorded should be detailed enough for others to reproduce the results of the experiment.

The experimental results are analyzed in depth, including the performance of the model, possible problems or room for improvement. In addition, the interpretability of the model needs to be analyzed in order to understand the decision-making process of the model.

3) MLOps model experiment process

MLOps pays more attention to the iteration speed and experimental efficiency of the model, shorens the experimental period and improves the experimental efficiency through automatic and intelligent methods. Automated machine learning and automated experimentation platforms are needed to quickly build and train models. The model needs to be evaluated and tuned to improve its performance and accuracy. Finally, the model needs to be deployed and applied to realize the practical application value of the model.

The model experiment design under MLOps pays more attention to iteration and efficiency. For example, in the same e-commerce recommendation system, an experimental design might be to quickly test different deep learning model architectures and use an automated platform for model training and evaluation.

Accelerate experiment execution with automated machine learning and automated experiment platforms. Automate end-to-end automation from data preparation to model training and evaluation with automated process management tools.

Rapid analysis and interpretation of experimental results for timely adjustment of model parameters or architecture. Focus on changes in assessment indicators to guide the design and execution of the next round of experiments.

4) Model evaluation

1.LLMOps Model Evaluation Process :

LLMOps prefers to use interpretable evaluation metrics such as accuracy, recall, etc., in order to better understand the performance of the model. For example, in an e-commerce recommendation system, LLMOps may focus on differences between different algorithms in the interpretation of recommendation results and user satisfaction.

2.MLOps model evaluation process:

MLOps focuses more on the use of complex and more representative metrics such as OC-ROC, F1 scores, etc., to assess model performance more comprehensively. In the e-commerce recommendation system, MLOps may pay attention to the overall performance of the recommendation system, including the comprehensive evaluation of accuracy, coverage, diversity and other indicators.

In conclusion, LLMOps and MLOps have some differences in the model experiment process and evaluation process, but both are committed to improving the performance and interpretability of the model to achieve better business results and user experience.

5) Algorithm selection and application

1.LLMOps algorithm selection and application:

LLMOps prefers to choose algorithms that have clear physical meaning and interpretability, such as decision trees, linear regression, etc., in order to better understand how the model operates. In addition, the stability and reliability of the algorithm need to be considered. The application process includes: algorithm selection and adjustment, model training and verification, model application and optimization.

2.MLOps algorithm selection and application:

MLOps pays more attention to the innovation and experimental effect of algorithms, and actively explores and applies new algorithms, such as neural networks and deep learning, to improve model performance. In addition, factors such as the advanced nature and efficiency of the algorithm need to be considered. The application process includes: algorithm research and selection, model design and training, model evaluation and application, etc.

6) Model deployment and monitoring

1.LLMOps model deployment and monitoring:

LLMOps emphasizes the stability and reliability of the model. In the process of model deployment and monitoring, it is necessary to formulate corresponding emergency plans, monitor the operation of the model in real time, and discover and solve problems in time. At the same time, LLMOps focuses on the interpretability and maintainability of the model, which helps to better understand the operation and performance of the model.

2.MLOps model deployment and monitoring:

MLOps focuses more on model efficiency and iteration speed, with intelligent methods to automate deployment and monitoring, reducing manual intervention. MLOps usually adopts real-time monitoring and alarm mechanism to discover and solve problems in time to ensure efficient and stable operation of the model. In addition, MLOps also focuses on the scalability and scalability of the model to adapt to different scenarios and requirements.

Through the above process, LLMOps and MLOps have different focuses on algorithm selection and application, model deployment and monitoring, but both aim to improve the performance, stability, and maintainability of models to meet business needs and improve user experience.

### 3.4 LLMOps application value

LLMOps is a method to deal with the future application of ultra-large scale machine learning technology to Ops operation and maintenance management, which can help enterprises more intuitively realize the automatic deployment and management of machine learning models through human natural language, and improve efficiency and reliability. Its application value is:

Automated machine learning model deployment: LLM's MLOps can reduce human intervention and improve model deployment efficiency and reliability by automating the deployment of machine learning models.

Model monitoring and management: LLM's MLOps can monitor and manage the running state of machine learning models, timely discover and solve problems in the model, and ensure the normal operation of the model.

Resource utilization optimization: LLMOps can monitor and analyze the resource utilization of IT systems through machine learning algorithms, optimize the resource allocation of the system, improve resource utilization, and reduce costs.

Automatic capacity expansion and contraction: LLMOps can automatically expand and shrink the capacity to cope with system load changes, allocate IT system resources based on actual requirements, and improve system performance and reliability.

Fault diagnosis and self-healing capability: LLMOps can diagnose and self-heal IT systems through machine learning algorithms, discover and solve faults in time, improve system reliability, and reduce downtime and maintenance costs.

Cost prediction and control: LLMOps can predict and control the operation and maintenance costs of IT systems through machine learning algorithms, helping enterprises better grasp the operation and maintenance costs, and optimizing the operation and maintenance strategies of future IT ultra-large scale algorithm model systems.

## 4. CONCLUSION AND PROSPECT

In conclusion, the advent of LLMOps heralds a new era in managing the lifecycle of LLM-driven applications, presenting both opportunities and challenges for enterprises. The complexity of engineering technology involved demands specialized teams for development and maintenance, while ensuring data security remains paramount to safeguard against breaches and misuse. Additionally, achieving model interpretability is crucial, especially in high-stakes applications like medical diagnosis or loan decisions, to ensure ethical and transparent decision-making processes.

Moreover, the integration of LLMOps into personalized recommendation systems underscores its potential in optimizing user experience. By leveraging LLMOps, enterprises can enhance the efficiency and reliability of large-scale machine learning models, driving personalized recommendations that align closely with user preferences. However, ethical and moral considerations must be addressed, particularly regarding the impact of AI systems on societal values and human welfare.

Looking ahead, LLMOps is poised to witness widespread adoption as AI technology continues to evolve. Enterprises will prioritize the management and maintenance of large-scale models, while also addressing concerns surrounding data privacy, security, and ethical implications. Ultimately, LLMOps will evolve to deliver more efficient, trustworthy, and secure machine learning services, ushering in a new era of personalized recommendation systems that elevate user experience to new heights.